# ROUTES TO BINARY GENE EXPRESSION

Indrani Bose

Department of Physics, Bose Institute, 93/1, A. P. C. Road, Calcutta-700009

___________________________________________________________________

*Systems biology approaches combining theoretical modeling with experiments have been singularly successful in uncovering novel features of cellular phenomena. One such feature is that of binary gene expression in which the expression level is either low or high, i.e., digital in nature. This gives rise to two distinct subpopulations in a population of genetically identical cells. The fraction of cells in the high expression state is raised as the strength of the inducing signal is increased indicating that the response is not graded. In this review, we discuss the possible origins of binary gene expression with emphasis on three principal mechanisms: purely stochastic, positive feedback-based and emergent bistability. In the latter case, two stable expression states are obtained due to an autoregulatory positive feedback loop in protein synthesis along with cell growth retardation by the proteins synthesized. The theoretical foundations of the observed phenomena are described in each case.*

___________________________________________________________________

## Introduction

Gene expression and its regulation are the most well-studied phenomena in cell biology. The subject, however, continues to yield surprises in the form of puzzling observations and novel features[1-4]. Systems biology approaches involving an interplay between theoretical modelling and quantitative measurements provide significant new insight on the origins and consequences of the novel features of gene expression. One such phenomenon, which has been observed in a large number of systems, is that of binary gene expression[5-8]. Consider a population of cells with identical genetic make-up and kept in the same environment. The expression level of a specific gene is expected to be more or less similar in all the cells. In the case of binary gene expression, the cell population develops into two distinct subpopulations. In one of the subpopulations, the protein level is low whereas in the other subpopulation the protein level is high. The fraction of cells in which proteins are present at intermediate levels is low. Binary gene expression has also been dubbed as the 'all-or-none' phenomenon the first evidence of which was provided in the pioneering experiment on the induction of the *lac* operon by Novick and Weiner[9]. When gene expression is regulated, an increase in the concentration of regulatory molecules (say, activators) raises the fraction of cells in the 'high' subpopulation resembling an analog-digital conversion. On the contrary, in the case of graded response, an increasing amount of stimulus yields a continuously changing response till a saturation level is reached.

Experimental observation of binary gene expression requires single cell measurements as population measurements yield only average values. Two well-known experimental techniques which provide information at the single cell level are: fluorescence microscopy and flow cytometry[10,11] both of which involve fluorescent reporter genes. The experimental observation of binary gene expression is in the form of a cell count versus fluorescence intensity distribution with two prominent peaks (figure 1). The smearing of the low and high protein levels into a two-peaked distribution is brought about by the noise associated with gene expression. Gene expression involves a series of biochemical events which are probabilistic in nature[1,2,12,13]. Due to the stochastic (random) nature of gene expression, the messenger RNA (mRNA) and protein levels in a cell exhibit fluctuations around a mean value as a function of time. In a population of cells, the protein levels in individual cells at a specific instant of time are not identical but spread around a mean value (figure 2). The mRNA/protein fluctuations around a mean level constitute noise for which appropriate quantification measures are available[1,2]. Simple models of stochastic gene expression, incorporating the essential aspects, have been developed the predictions of which have been tested in actual experiments[1,2]. Single cell and single molecule techniques for the study of gene expression show that the mRNAs and proteins are produced in abrupt stochastic bursts[12,13]. This contrasts with the textbook picture that transcription and translation proceed smoothly and uniformly[14]. Noise, far from being just a nuisance, plays an essential role in a number of cellular processes, e.g., cell differentiation, cellular decision making and generation of phenotypic heterogeneity in microorganisms as a survival strategy[1,2,15-19]. In the following, we review the major routes to binary gene expression in all of which the transitions between the two expression states are noise-driven. We first discuss the situation in which stochasticity in gene expression by itself is sufficient to generate a bimodal distribution in the protein levels. We next describe the role of positive feedback in genetic circuits exhibiting binary gene expression. We finally point out a recent example[20] of binary gene expression, termed 'emergent bistability', which illustrates a hitherto-unknown mechanism for the generation of two-peaked protein distributions.

## Stochastic Binary Gene Expression

In a simple stochastic model of gene expression[21-23], a gene can be in two possible states: inactive (G) and active (G*). Protein synthesis can occur only in the active state of the gene whereas protein decay occurs in both the states. Transitions between the states G and G* are stochastic in nature, i.e., occur at random time intervals. The reaction scheme describing the above mentioned processes is given by

$$G \xrightarrow{k_a} G^*$$

$$G^* \xrightarrow{k_d} G$$

$$G^* \xrightarrow{j_p} p$$

$$p \xrightarrow{k_p} \phi \qquad (1)$$

where $k_a$ and $k_d$ are the activation and deactivation rate constants respectively, $j_p$ is the rate constant for protein synthesis and $k_p$ the rate constant for protein decay. The protein decay rate has two components: the degradation rate and the dilution rate due to cell growth and division. In the case of stochastic gene expression, the rate constants have a probabilistic interpretation. For example, $k_a dt$ gives the probability that the gene switches to the active state $G^*$ in the time interval $(t + dt)$ given the gene was in the inactive state G at time t. While the full stochastic model is analytically tractable[21,24], Karmakar and Bose[8] considered a simpler version of the model[22] in which the stochasticity is associated with only the random transitions between the two gene states, G and $G^*$. Protein synthesis and protein decay occur in a deterministic manner. In each state of the gene, the concentration of proteins evolves deterministically according to the equation

$$\frac{dx}{dt} = \frac{j_p}{X_{max}} z - k_p x \qquad (2)$$

where $z = 1$ (0) in the active (inactive) state of the gene and $x = X/X_{max}$, X and $X_{max}$ ( $= j_p / k_p$ ) being the protein concentration at time t and the maximum protein concentration respectively. The value of z flips between '0' and '1' at random time intervals. The probability density function describing the distribution of protein levels in the steady state is given by the beta distribution[8]

$$p(x) = N x^{\left(\frac{k_a}{k_p} - 1\right)} (1-x)^{\left(\frac{k_d}{k_p} - 1\right)} \qquad (3)$$

where $N = \frac{\Gamma(r_1 + r_2)}{\Gamma(r_1)\Gamma(r_2)}$ with $r_1 = k_a / k_p$ and $r_2 = k_d / k_p$. In the expression for N, $\Gamma(y)$ represents the well-known gamma function. Also, the fraction of cells in which the protein level lies between x and x + dx is given by p(x)dx. In the case of inducible gene expression, activators or transcription factors (TFs) promote transitions to the active state $G^*$ of the gene. In this case, $k_a$ and $k_d$ are functions of the concentration of the activator/TF molecules. Considering the parameters $r_1$ and $r_2$ in Eqn. (3), four cases are to be considered separately: (i) $r_1 < 1$, $r_2 < 1$, (ii) $r_1 > 1$, $r_2 > 1$, (iii) $r_1 < 1$, $r_2 > 1$ and (iv) $r_2 < 1$, $r_2 > 1$. In the first case, binary gene expression in the form of a bimodal distribution in the protein levels is obtained (figure 3). In the case of inducible gene expression, as the concentration S of the activator/TF molecules is changed, one obtains a binary response, i.e., as S increases the fraction of cells in the 'high' subpopulation goes up. In the second case, a unimodal (single peak) distribution with graded response is observed. The peak position shifts as a function of S. In the third and fourth cases, unimodal non-graded responses are obtained.

Let us now discuss the physical origin of stochastic binary gene expression. If the gene is always in the inactive state ($z = 0$ in Eqn. (2)), the mean protein level in the steady state is given by $x = 0$. If the gene is always in the active state ($z = 1$ in Eqn. (2)), the steady state protein level is $j_p/k_p$ so that $x = 1$ corresponding to maximal protein synthesis. When random activation and deactivation processes are taken into account, two major possibilities arise. If the activation and deactivation rates are faster than the protein decay rate ($r_1 > 1$, $r_2 > 1$), an average protein level, intermediate between $x = 0$ and $x = 1$, is obtained due to the accumulation of proteins over the random transitions between the inactive and active states of the gene. In the opposite case, i.e., when the activation and deactivation rates are slower than the protein decay rate ($r_1 < 1$, $r_2 < 1$), the mean protein level is either $x = 0$ or $x = 1$ depending on whether the gene is in the inactive or the active state. This is because a sufficient time is available in the active state for the protein to reach the steady state level $x = 1$. The residence time of the gene in the inactive state is sufficiently long so that the accumulated proteins decay fully during the time interval of expression inactivity resulting in the attainment of the steady state level $x = 0$.

There are now several examples of stochastic binary gene expression in eukaryotes brought about by slow promoter transition rates[1,2,8]. Karmakar and Bose[8] showed that the simple stochastic model of gene expression considered by them can fit the experimental data of Zlokarnik et al.[25] on binary gene expression quite well. In the model considered, one could replace protein synthesis and decay by mRNA synthesis and decay to obtain the steady state distribution of mRNA levels. Raj et al.[26] in their experiment on stochastic mRNA synthesis in Chinese hamster ovary cells found evidence that the dominant stochasticity is associated with the random transitions between the inactive and active states of the gene. The mRNA molecules are synthesized during intense 'on' periods (transcriptional bursts) while the 'off' periods are of longer duration. In the case when the rate of inactivation, $k_d$, is significantly larger than the activation rate, $k_a$, and is larger than the mRNA decay constant ($k_p$ in Eqn. (2) now represents the mRNA decay rate constant), the beta distribution (Eqn. (3)) reduces to the gamma distribution[26,27]. Slow transitions between the inactive and active promoter (gene) states are expected to be relevant in eukaryotic gene expression[1,2]. Transition between closed and open chromatin (containing DNA-nucleosome complexes) structures is required for the promoter to be accessible to the transcriptional machinery. Chromatin remodeling, it is known, can be quite slow[1,2], also the appropriate assembly of the TFs and the RNAP complex at the specific region of the DNA takes a certain amount of time to be completed. Once transcription is initiated, the gene continues to be in the active state as long as the transcriptional complexes are all in place. Chance unbindings of one or more components bring about a transition to the inactive state of the gene. Figure 4 shows how binary gene expression (two distinct subpopulations) originates due to slow transitions between the inactive and active states of the gene.

## Positive Feedback-based Binary Gene Expression

Positive feedback loops constitute one of the most common motifs in gene transcription regulatory networks and signaling cascades. The simplest such motif is an autoregulatory loop in which the proteins synthesized by a gene function as regulatory molecules to promote their own synthesis. Positive feedback combined with sufficient nonlinearity in the gene expression dynamics may give rise to bistability in a range of parameter values[6,28,31]. Bistability implies that the cell has a choice between two stable expression states for the same parameter values. The choice of the stable steady state is dictated by the previous history of the system indicative of cellular memory[32]. We illustrate the physical origin of bistability by treating the example of the autoregulatory positive feedback loop discussed above. Let X represents a protein molecule synthesized by gene G (figure 5(a)). The protein binds the promoter region of the gene to activate the initiation of transcription. The protein molecule may also unbind as shown in figure 5(a). The rate of change of the protein concentration x is given by

$$\frac{dx}{dt} f(x) - g(x) \qquad (4)$$

where f(x) and g(x) represent the rates of increase and decrease respectively of the protein concentration. We first derive the functional form of f(x). Let $k_f$ and $k_d$ be the binding and unbinding rate constants for the protein-DNA complex. The equilibrium condition is achieved when the rate of binding equals the rate of unbinding, i.e.,

$$k_f \, x \, [G] = k_b \, [xG] \qquad (5)$$

where [G] and [xG] denote the concentrations of genes without and with bound protein respectively. If [$G_{tot}$] is the total concentration of genes, then

$$[G_{tot}] = [G] + [xG] \qquad (6)$$

One then gets

$$[xG] = \frac{G_{tot} \, x}{(K_d + x))} \qquad (7)$$

where $K_d = k_b / k_f$ is the equilibrium dissociation constant. Since, transcription is initiated only when the protein X binds the gene, the probability that a gene is ready to be transcribed, i.e., in the active state is given by [xG]/[$G_{tot}$]. If $j_p$ is the rate constant for protein synthesis, the rate of increase in protein concentration, due to protein synthesis, is given by

$$f(x) = \frac{j_p \, x}{(K_d + x)} \qquad (8)$$

The rate of decrease in protein concentration, due to protein decay, is given by

$$g(x) = k_p\, x \qquad (9)$$

Figure 5(b) shows the plots of f(x) and g(x) versus x. At the points of intersection of the two curves, f(x) = g(x) so that dx/dt = 0 (Eqn. (4)). The values of x, $x_1$ and $x_2$, at the points of intersection define the steady states of the system. A steady state is said to be stable (unstable) if the system regains (does not regain) the steady state after being weakly perturbed from it. A simple graphical analysis shows that $x_1$ ($x_2$) defines an unstable (stable) steady state. Consider the steady state corresponding to $x_2$. Let the new value of x be to the right of $x_2$ (figure 4(b)) due to a weak perturbation. In this case, f(x) is < g(x) so that dx/dt is < 0. Thus as time increases, shown by an arrow direction in figure 5(b), x decreases. Once x reaches the value $x_2$, further time evolution stops. If x is originally to the left of $x_2$, a similar argument shows that the system regains the steady state corresponding to $x_2$ which is thus a stable steady state. One can further show that the steady state defined by $x_1$ is unstable.

Figure 6(a) shows the situation when two protein molecules, represented by X, form a dimer. The dimer binds the promoter region of the gene G to activate the initiation of transcription. The dimer may unbind as well as dissociate into two single proteins. The rate of change of protein concentration is given by Eqn. (4) with

$$f(x) = j_p \frac{x^2}{K_d^2 + x^2}$$

$$g(x) = k_p\, x \qquad (10)$$

where $K_d^2$ is now $K_d^2 = (k_b\, k_{-1})/(k_f\, k_1)$; $k_1$ and $k_{-1}$ are the dimer formation and dissociation constants respectively. Figure 6(b) shows that in a specific parameter regime, three steady state solutions $x_1$, $x_c$ and $x_2$ are possible with $x_1$ and $x_2$ corresponding to stable steady states and $x_c$ defining an unstable steady state. This is the well-known case of bistability which turns out to be a universal theme in several cell biological processes[28,33,34,35]. In the example under consideration, the three steady state solutions (dx/dt = 0 in Eqn. (4)) are: $x_1 = 0$,

$x_c = \dfrac{\left(j_p - \sqrt{j_p^2 - 4k_p^2\, K_d^2}\right)}{2k_p}$ and $x_2 = \dfrac{\left(j_p + \sqrt{j_p^2 - 4k_p^2\, K_d^2}\right)}{2k_p}$. Valid physical solutions are obtained in the parameter regimes for which $x_c$ and $x_2$ are > 0. The two solutions $x_1$ and $x_2$ coalesce when $j_p = 2k_p^*\, K_d$ with $x_c = x_2 = K_d$, $k_p^*$ denotes the special value of the decay rate constant $k_p$ at which the two steady state solutions merge. For $k_p > k_p^*$, the system is monostable with only one stable steady state at $x_1 = 0$. For $k_p < k_p^*$, the autoregulatory positive feedback module exhibits bistability. Thus, changing a parameter of the system, namely, $k_p$, one can bring about a transition from monostability to bistability.

In general, bistability is often accompanied by the interesting feature of hysteresis[19,30,31,34,35]. Figure 7(a) illustrates the origin of hysteresis in the case of a regulatory signal evoking a steady state response. The shaded region represents the region of bistability. The solid lines indicate stable steady states whereas the dotted line represents the branch of unstable steady states. Let us suppose that the stable steady state of the system is initially on the lower branch. As the regulatory signal increases in strength, the steady state continues to be on the lower branch till a 'bifurcation' point is reached. At this point there is a discontinuous jump to the upper branch of stable steady states. If one now reverses the direction of change in the strength of the regulatory signal, one finds that the change in the response is not reversible. There is a downward jump from the upper to the lower branch only at a lower 'bifurcation' point. This type of behavior is designated as hysteresis. The presence of hysteresis explains the experimental observation by Novick and Weiner[9] that the strength of the inducing signal required to maintain the induced state is lower than that needed to switch from the uninduced to the induced state. In the region of bistability, the choice of a stable steady state is dictated by history, i.e., past events. The magnitude of the protein concentration in the unstable steady state sets a threshold. In the example of the autoregulatory positive feedback module, $x_c$, for example, sets the threshold. If the initial protein concentration x lies in the range $0 < x < x_c$, the system evolves to the steady state defined by $x_1$. If $x_c < x$, the steady state corresponding to $x_2$ is reached in the course of time. If the cells in a population are in the same initial state, the steady state should be the same in each cell. How does then population heterogeneity in the form of two distinct subpopulations occur? Again, the stochastic nature of gene expression has to be taken into account to explain the heterogeneity. A transition from the low to the high expression state is brought about once the fluctuations associated with the low expression level cross the threshold set by the unstable steady state (figure 7(b)). Noise-induced transitions give rise to a bimodal distribution in the protein levels (figure 7(c)).

A number of examples is now known[28,30,33,34,36] in which binary gene expression based on bistability has been observed experimentally. In some recent experiments[37,38], bistability accompanied by hysteresis has been demonstrated in a population of M. smegmatis subjected to nutrient depletion as a source of stress. The distributions of key regulatory proteins in the stress response pathway have been measured in flow cytometry experiments and in each case the nature of the distributions is found to be bimodal. The results of a theoretical model along with a comprehensive analysis of flow cytometry data provide definitive evidence that the distribution of GFP (green fluorescent protein acting as the reporter) levels at any point of time along the growth curve is a linear superposition of two invariant distributions, one Gaussian and the other lognormal, with only the coefficients in the linear combination depending on time[38].

A common survival strategy of microorganisms subjected to stress involves the generation of phenotypic heterogeneity in the isogenic microbial population, enabling a subset of the population to survive under stress. Some prominent examples involving positive feedback include: lysis/lysogeny in bacteriophage λ[39], competence development in B.subtilis[28,40,41,42], infection of a host cell by HIV-1 virus with a subsequent choice between two distinct cell fates, latency and lysis[43] and the stringent response in mycobacteria[37,38]. In the case of B. subtilis, the protein ComK plays the key role in the development of competence in a fraction of cells. Binary gene expression, in the form of low and high ComK levels, may be an outcome of bistability due to the presence of a positive feedback loop involving ComK proteins activating their own synthesis or may be a result of excitability due to the coupled dynamics of a fast positive and a slow negative feedback loop[40]. The role of gene expression noise in bringing about the switch to competence is, however, well demonstrated experimentally[41]. In the case of HIV-1 virus, the key regulatory protein is Tat and the relevant circuit dynamics give rise to a long lived pulse of Tat proteins[43]. High levels of Tat bring about cell lysis.

## Emergent Bistability

In general, positive feedback and cooperativity in the regulation of gene expression are necessary to obtain bistability. Recently, Tan et al.[20] have proposed a new mechanism by which a noncooperative positive feedback circuit and circuit-induced growth retardation of the cell give rise to bistability. The novel type of bistability was demonstrated in a synthetic gene circuit. The circuit consists of a single autoregulatory positive feedback loop in which the protein product X of a gene promotes its own synthesis in a noncooperative fashion. As mentioned earlier, the protein decay rate has two components, the degradation rate and the dilution rate due to cell growth. In the synthetic circuit under consideration, production of X slows down cell growth so that the rate of dilution of X and consequently the protein decay rate are reduced. This generates an effective positive feedback loop since an increased synthesis of X proteins leads to a greater accumulation of the proteins which in turn promote further protein synthesis. The combination of the effective positive feedback loop and the original positive feedback loop give rise to bistability in the absence of cooperativity. A related study by Klumpp et al.[44] has also considered a positive feedback loop generated due to the cell growth inhibition by a protein. The dynamics of the synthetic gene circuit can be described by the following equation[20]

$$\frac{dx}{dt} = \frac{\alpha_0 + \alpha_m x}{K_d + x} - \frac{\mu_{max} x}{1 + \theta x} - k_p x \qquad (11)$$

where x denotes the protein concentration, $\alpha_0$ is the basal rate of protein synthesis and $\alpha_m$ the rate of regulated protein synthesis, $K_d$ is the dissociation constant for the unbinding of proteins from the promoter region of the gene, $\mu_{max}$ is the maximum growth rate, $\theta$ is a parameter representing the 'metabolic burden' (reduced cell growth rate due to protein synthesis) and $k_p$ is the decay rate constant of the proteins. We specifically note the presence of the nonlinear term in Eqn. (11) representing along with the protein decay term the net rate of decrease in the protein concentration. The first term in Eqn. (11) represents f(x) in Eqn. (4) and the last two terms together constitute g(x) (Eqn. (4)). Figure 8 illustrates the origin of bistability in the gene circuit. In the absence of protein induced cell growth retardation, one obtains monostability as there is only one point of intersection of the f(x) and g(x) versus x curves. In the presence of the second positive feedback loop, bistability is possible when the f(x) and g(x) curves intersect at three points. The OFF and ON states are the stable steady states separated by an unstable steady state. A recent study[38] indicates that the mechanism of emergent bistability, based on protein induced cell growth retardation, may explain the observed binary gene expression in mycobacteria subjected to stress.

## Concluding Remarks

A large number of studies, both theoretical and experimental, have been carried out in the last few years on the origins and consequences of binary gene expression. In this review, we have discussed three mechanisms via which two expression states are achieved. In all the cases, gene expression noise plays a key role in bringing about transitions between the low and high expression states. A recent experiment[45] illustrates how variations of the broad themes outlined in this review are possible. Using a synthetic system in budding yeast, To and Maheshri[45] showed that positive feedback involving a promoter with multiple TF binding sites can exhibit binary gene expression without cooperative binding of the TFs. The bistability is not predicted by deterministic models but is an outcome of a short-lived TF and stochastic fluctuations in the TF's expression.

As discussed earlier, bistability is often accompanied by hysteresis which enhances the robust functioning of the gene circuit by making reverse transitions difficult once a switch is made to the induced/high expression state. If the expression of a target gene exhibits ultrasensitivity as a function of changing stimulus strength, one can define a threshold below which the response (target gene expression level) is low and above which the response is high. Again, fluctuations in the stimulus amount close to the threshold can give rise to two distinct subpopulations corresponding to low and high response respectively. There are a number of known examples in which this feature has been observed experimentally[1,2,46]. In the case of an excitable dynamical system[40,42,43], there is only one stable steady state and two unstable steady states. Fluctuations in the protein level associated with the stable steady state can activate a switch to expression states in the vicinity of one of the unstable steady states. The transient activation is followed by a return to the original stable expression state. In this case also, two subpopulations with distinct gene expression levels are observed. We have not discussed such cases in detail in the review.

Microorganisms adopt a number of strategies to cope with stressful conditions like environmental fluctuations, nutrient depletion and application of antibiotic drugs[16,17,19]. One such strategy is the creation of phenotypic heterogeneity so that the whole population does not suffer the same fate. Positive feedback combined with noise-induced transitions between the expression states constitute one mechanism for the generation of phenotypic heterogeneity. Pathogens like mycobacteria exhibit a remarkable ability to survive under stress; a fraction of the mycobacterial population (the so-called persisters) survives in the lung granulomas in which there is a paucity in life-sustaining agents. The persister subpopulation owes its origin to positive feedback and stochastic gene expression[37,38]. The concentrations of key regulatory proteins in the stress response pathway, with nutrient-depletion serving as the source of stress, are high in the persister sub-population and low in the rest of the population. Though binary gene expression has been observed in the mycobacterial species M.smegmatis[37,38], the same type of stress response is expected in M. tuberculosis, the pathogen causing tuberculosis. Thus binary gene expression may provide a basis for developing effective drug strategies which act against the generation of phenotypic heterogeneity.

Figure Captions

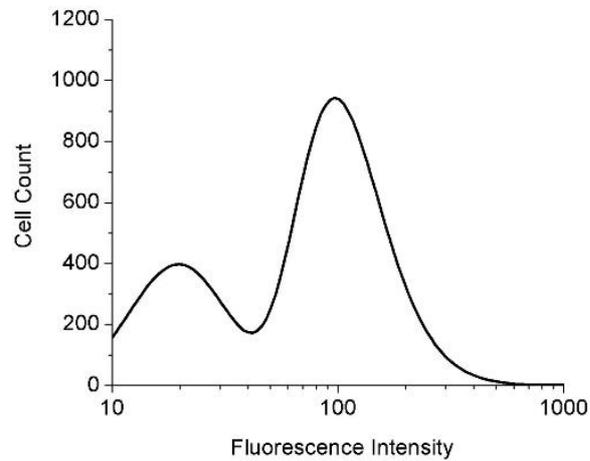

**Figure 1.** Binary gene expression gives rise to a bimodal distribution of protein levels in a population of cells.

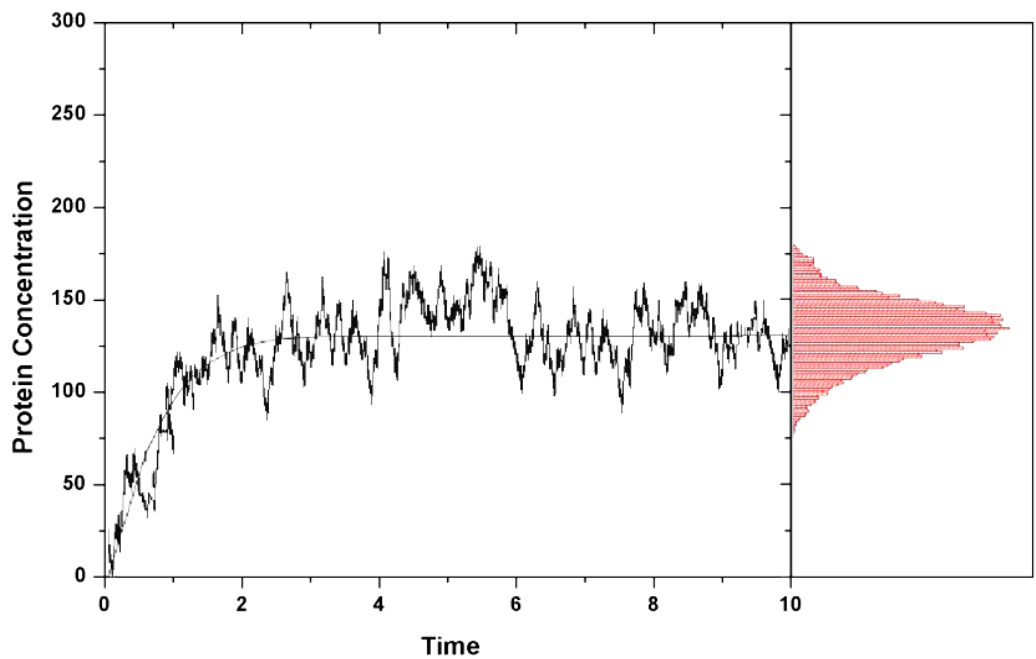

**Figure 2.** Protein concentration as a function of time exhibits fluctuations around the mean level. The histogram shows the spread of steady state protein levels in a population of cells.

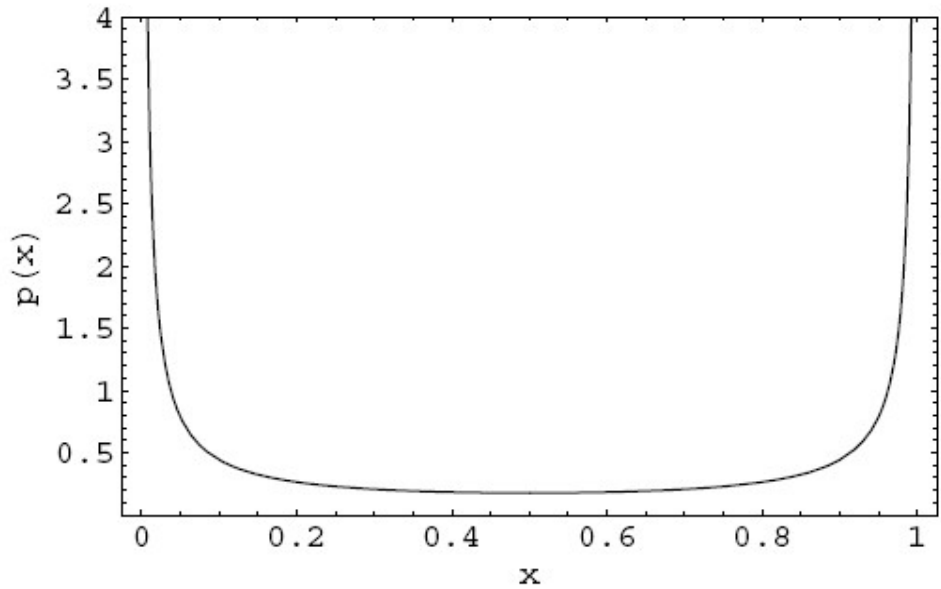

**Figure 3.** Stochastic binary gene expression described by the beta distribution (Eqn. (3)) for $r_1 < 1$ and $r_2 < 1$.

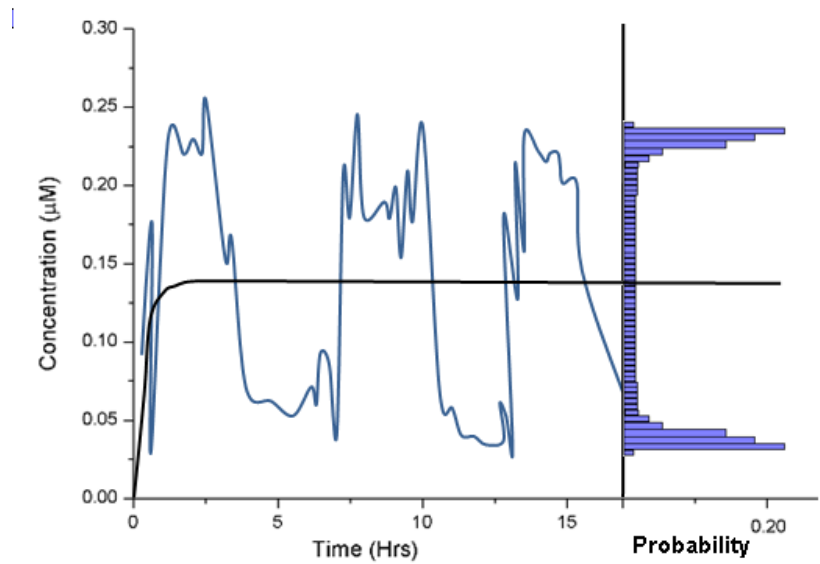

**Figure 4.** Binary gene expression due to slow transitions between the inactive and active states of the gene.

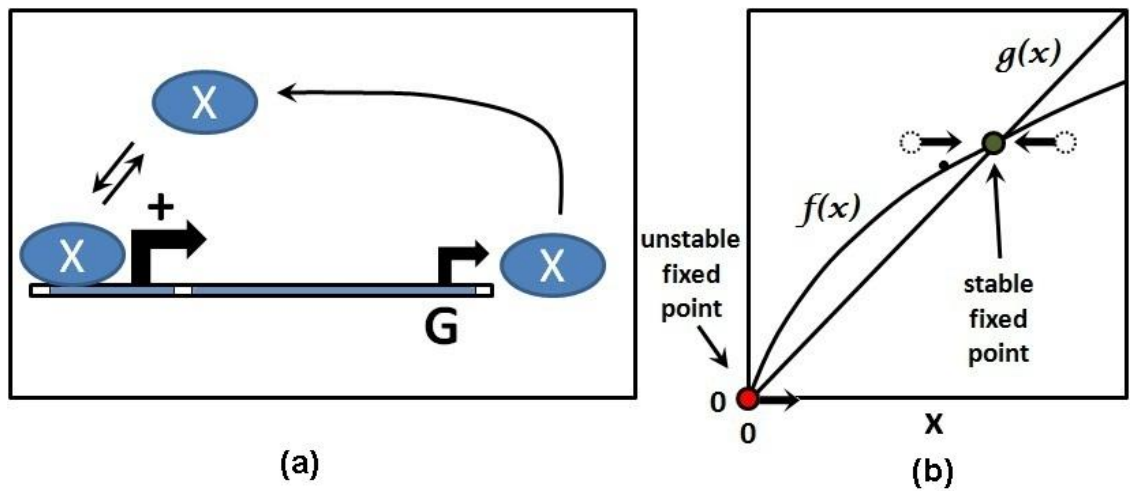

**Figure 5.** (a) An autoregulatory positive feedback module in which a protein X promotes its own synthesis. (b) The plots f(x) and g(x) (Eqns. (8) and (9)) versus x intersect at two points representing two steady states. Only one of the steady states is stable.

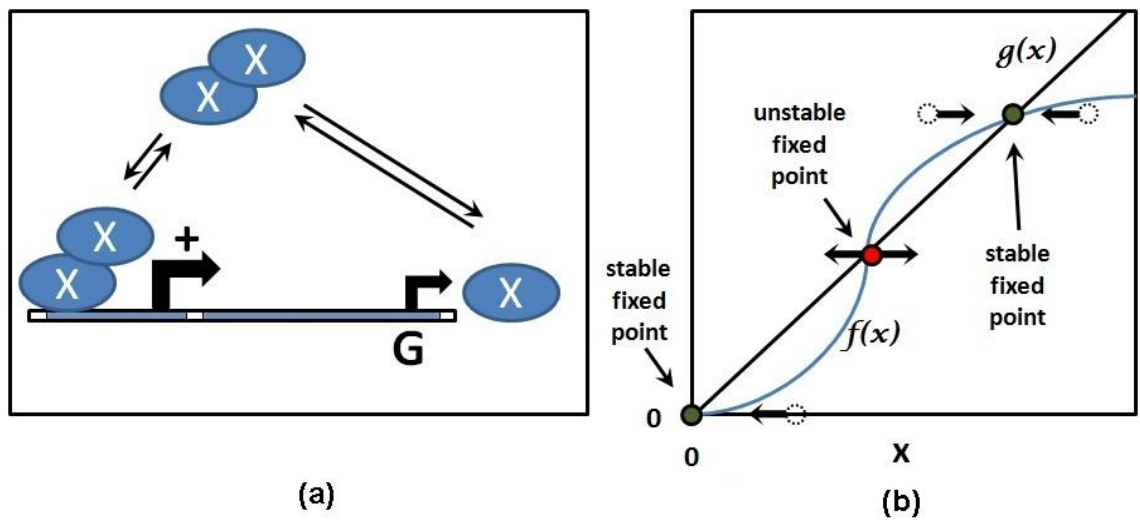

**Figure 6.** (a) Two protein molecules, represented by X, form a dimer which, on binding the promoter region, activates the initiation of transcription. (b) The plots f(x) and g(x) (Eqn. (10)) versus x intersect at three points. The two stable steady states are separated by an unstable steady state.

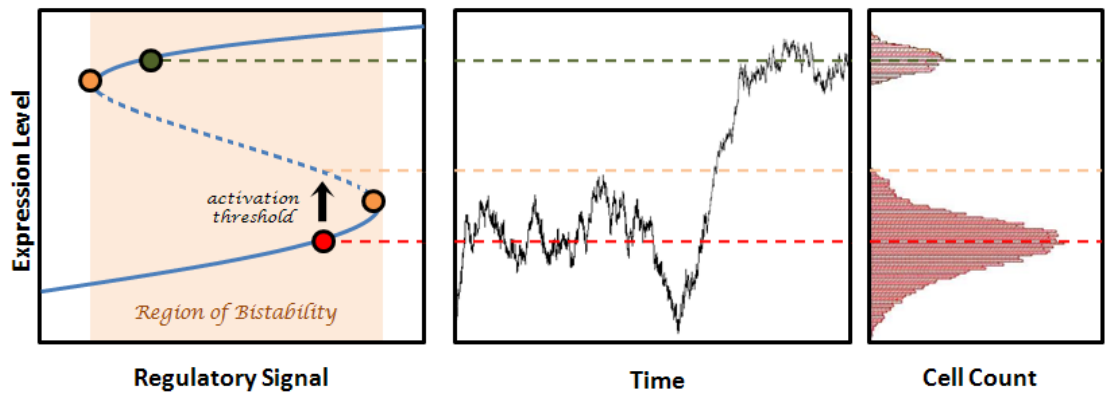

**Figure 7.** (Left panel) Steady state expression level versus regulatory signal amount exhibits hysteresis. The solid lines represent stable steady states. The shaded region corresponds to bistability with the points of inflection denoting the bifurcation points. The middle panel shows noise-induced transition from the low to the high expression state. The right panel exhibits a bimodal distribution of protein levels in a population of cells.

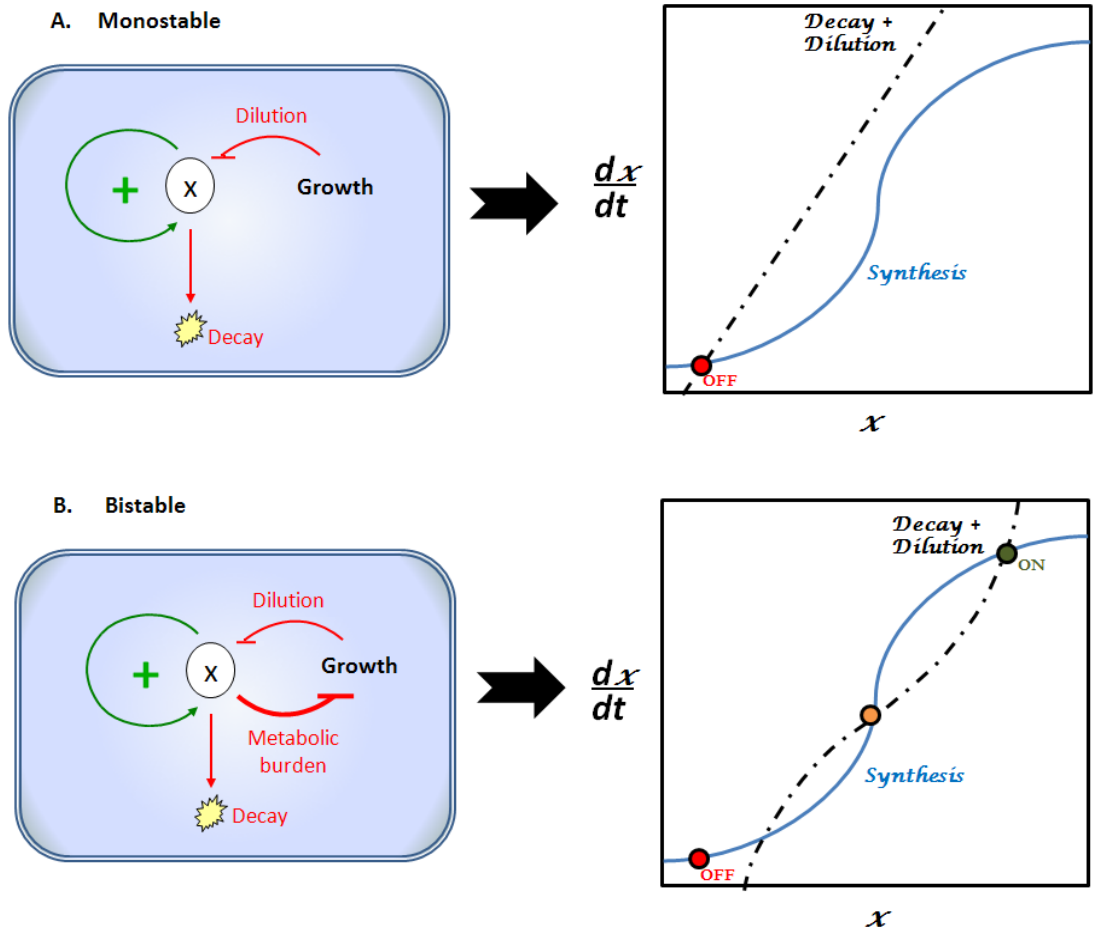

**Figure 8.** A. An autoregulatory positive feedback loop without multimerization of the protein molecules gives rise to one stable steady state. B. Protein induced cell growth retardation gives rise to a second positive feedback loop. In this case, bistability is obtained corresponding to low (OFF) and high (ON) gene expression states.